\documentclass[%
 reprint,
 superscriptaddress,
 nofootinbib,
 amsmath,amssymb,
 prb,
 floatfix,
]{revtex4-1}

\usepackage[utf8]{inputenc}
\usepackage[T1]{fontenc} 

\usepackage{graphicx}
\usepackage{dcolumn}
\usepackage{bm}
\usepackage{braket}
\usepackage{color}
\usepackage{multirow}
\usepackage[percent]{overpic}
\usepackage{changes}
\usepackage{bbold}
\usepackage{amsmath}
\usepackage{changes}

\renewcommand{\t}{\text}
\newcommand{\brak}[1]{\left( #1 \right)}
\newcommand{\abs}[1]{\left| #1 \right|}

\newcommand{\braq}[1]{\left[ #1 \right]}
\newcommand{\e}{\mathrm{e}}
\renewcommand{\vec}[1]{\textbf{#1}}

\begin{document}

\title{Coulomb-Engineered Heterojunctions and Dynamical Screening in Transition Metal Dichalcogenide Monolayers}

\author{C. Steinke}
	\affiliation{Institut f{\"u}r Theoretische Physik, Universit{\"a}t Bremen, Otto-Hahn-Allee 1, 28359 Bremen, Germany}
	\affiliation{Bremen Center for Computational Materials Science, Universit{\"a}t Bremen, Am Fallturm 1a, 28359 Bremen, Germany}
	
\author{T. O. Wehling}
	\affiliation{Institut f{\"u}r Theoretische Physik, Universit{\"a}t Bremen, Otto-Hahn-Allee 1, 28359 Bremen, Germany}
	\affiliation{Bremen Center for Computational Materials Science, Universit{\"a}t Bremen, Am Fallturm 1a, 28359 Bremen, Germany}
	
\author{M. R\"osner}
	\email{m.roesner@science.ru.nl}
	\affiliation{Institute for Molecules and Materials, Radboud University, Heijendaalseweg  135, 6525 AJ Nijmegen, The Netherlands}

\date{\today}

\begin{abstract}

The manipulation of two-dimensional materials via their dielectric environment offers novel opportunities to control electronic as well as optical properties and allows to imprint nanostructures in a non-invasive way.
Here we asses the potential of monolayer semiconducting transition metal dichalcogenides (TMDCs) for Coulomb engineering in a material realistic and quantitative manner. 
We compare the response of different TMDC materials to modifications of their dielectric surrounding, analyze effects of dynamic substrate screening, i.e. frequency dependencies in the dielectric functions, and discuss inherent length scales of Coulomb-engineered heterojunctions. 
We find symmetric and rigid-shift-like quasi-particle band-gap modulations for both, instantaneous and dynamic substrate screening.
From this we derive short-ranged self energies for an effective multi-scale modeling
of Coulomb engineered heterojunctions composed of an homogeneous monolayer placed on a spatially structured substrate.
For these heterojunctions, we show that band gap modulations on the length scale of a few lattice constants are possible rendering external limitations of the substrate structuring more important than internal effects.
We find that all semiconducting TMDCs are similarly well suited for these external and non-invasive modifications. 
\end{abstract}


\maketitle

\section{Introduction}

\begin{figure}
		\includegraphics[width=0.9\columnwidth]{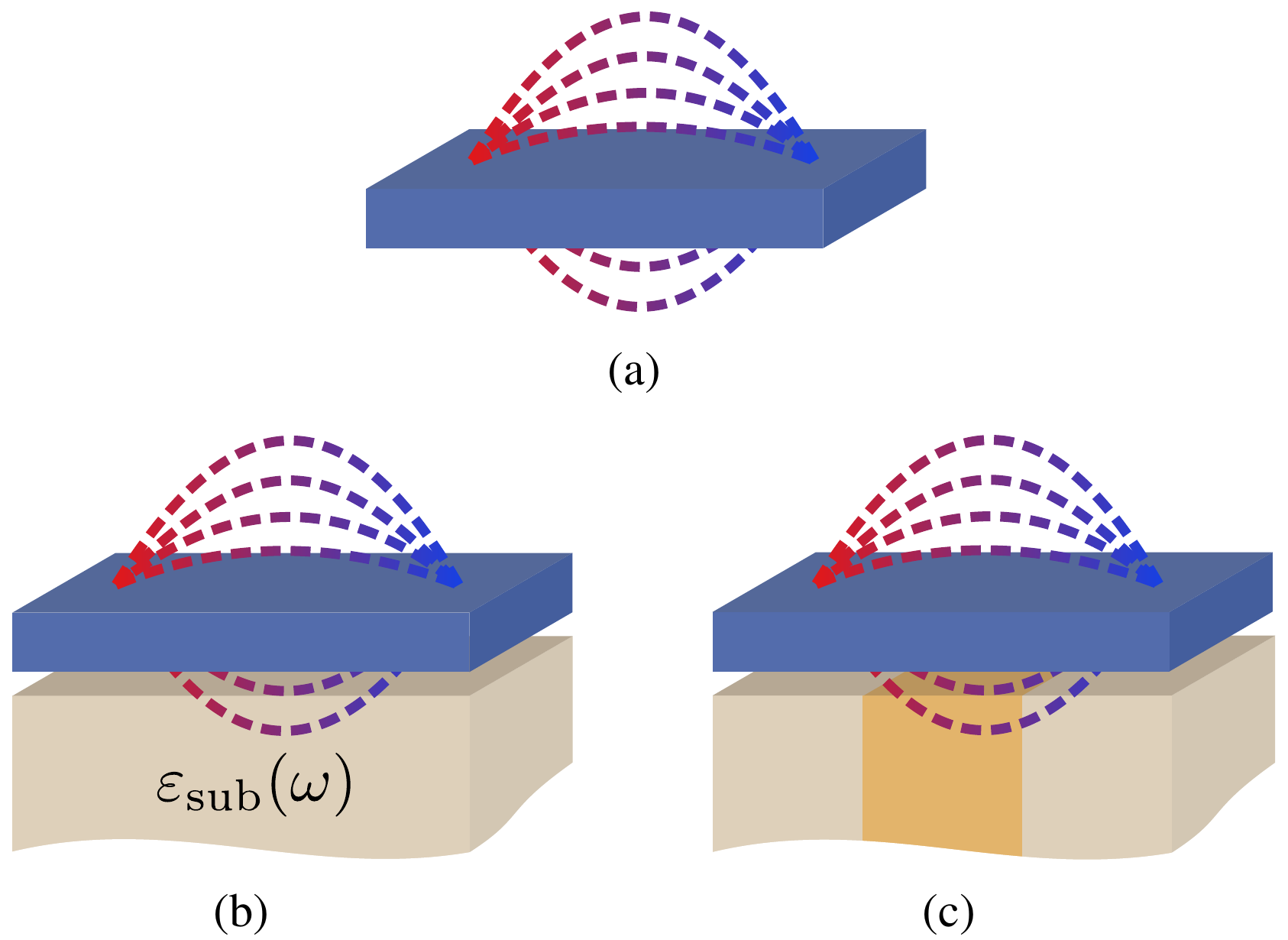}
		\caption{Sketches of (a) a free-standing monolayer, (b) a monolayer on a dielectric substrate, and (c) a Coulomb-engineered heterostructure. \label{fig-1} }
	\end{figure}

	In (quasi) two-dimensional (2D) materials the Coulomb interaction is enhanced due to weak intrinsic screening \cite{Keldysh1979,Cheiwchanchamnangij2012,Ramasubramaniam2012,Berkelbach2013}. Modifications of the immediate surrounding via substrates, capping layers or adsorbates as depicted in Fig.~\ref{fig-1} can therefore strongly affect the Coulomb interaction and its related effects.
	As a result the band gaps of 2D semiconductors are, for example, strongly influenced by the chosen substrate or capping material\cite{komsa_effects_2012, Hser2013-2, Qiu2013, Ugeda2014, Park_2018}. 
	By embedding 2D materials in spatially inhomogeneous dielectric environments,
	\textit{Coulomb engineered} heterostructures with spatially changing quasi-particle band gaps \cite{rosner_two-dimensional_2016,ryou_monolayer_2016,raja_coulomb_2017,Borghardt2017,Utama2019} can be created. In recent years, this approach to non-invasively manipulate 2D materials has become a promising field of research \cite{ryou_monolayer_2016,bradley_probing_2015,zhang_bandgap_2016,stier_probing_2016,Florian_dielectric_2018,Trolle2017,Latini2015,Bruix2016,Winther2017,Cho2018,Waldecker2019,Steinhoff2018,Jia2019}. 
	
	Here, we extend our previous studies\cite{Waldecker2019,Steinhoff2018,rosner_two-dimensional_2016}, present a detailed investigation of Coulomb engineering effects to semiconducting transition metal dichalcogenides (TMDCs), and analyze the influence of frequency dependencies in the substrate dielectric functions on the band structure. Since full ab-initio $GW$ calculations of lateral or vertical 2D heterostructures are numerically very demanding, we implement a description based on a combination of the $G \Delta W$\cite{rohlfing_electronic_2010,winther_band_2017} and Wannier Function Continuum Electrostatics (WFCE)\cite{rosner_wannier_2015} approaches going beyond our previous Hartree-Fock model study from Ref.~\onlinecite{rosner_two-dimensional_2016}. In this way we systematically investigate semiconducting TMDCs in their H-phase and present a material realistic, i.e. ab-initio based, modeling scheme to describe Coulomb-engineered 2D material based systems.
	
	For homogeneous substrates, as depicted in Fig.~\ref{fig-1} (b), we find that the effects of instantaneous (frequency independent) dielectric functions \footnote{If screening is instantaneous, then the dielectric function is frequency independent and the static dielectric function $\varepsilon (\omega = 0)$ is applicable at all frequencies. This is why, the term \textit{static screening} is sometimes intermixed with the term \textit{instantaneous.}}  are comparable for all materials under investigation with the transition metal sulfides being slightly stronger affected. For retarded (frequency-dependent) substrate screening  $\varepsilon_\text{sub}(\omega)$ we find symmetric shifts of the valence and conduction bands with slightly enhanced effects in the sulfides. 
	We show that these effects from retarded dielectric functions can be approximately mapped to effective static dielectric constants.
	We furthermore find spatially strongly localized self-energies, which are resulting from material-intrinsic properties independently of the surrounding material. This is again very similar for all TMDCs under investigation. Based on these major findings, we can subsequently construct an efficient approach to describe Coulomb-induced heterojunctions as depicted in Fig.~\ref{fig-1} (c) on an ab-initio basis.
	
	The paper is structured as follows: in section \ref{sec:method}, we introduce our theoretical approach. In the results section \ref{sec:results}, we discuss in detail the influence of homogeneous substrates with instantaneous (sec. \ref{sec:results_static}) as well as retarded dielectric functions (sec.~\ref{sec:results_dynamic}) on monolayer TMDCs. We analyze in section \ref{sec:results_length} the substrate-induced self-energy corrections, which we show to be short ranged for all materials under investigation. On this basis, a multi-scale approach for the simulation of Coulomb interaction effects in dielectrically engineered heterostructures is laid out and applied to the example case of WS$_2$ in section \ref{sec:results_heterostruct}.
	
\section{Method} \label{sec:method}

	The band structure of any solid-state material results from single particle contributions and is influenced by the many-body  Coulomb interaction $W$. 
	In the following we aim to understand how changes to $W$ of a monolayer TMDC induced by modifications of its dielectric environment affect its electronic quasi-particle band structure and particularly its band gap. 

	The screened Coulomb interaction $W$ is in general non-local and frequency-dependent. For a freestanding TMDC monolayer it reads in momentum space:
	\begin{align}
		W^\text{TMDC}(\vec{q}, \omega) 
			= \frac{v(\vec{q})}{ \varepsilon^\text{TMDC}(\vec{q}, \omega) } \label{eq:coulomb_FS},
	\end{align}
	where $v(\vec{q})$ is the bare Coulomb interaction, $\varepsilon^\text{TMDC}(\vec{q}, \omega)$ the dielectric function of the monolayer including only internal screening effects, and $\vec{q}$ the in-plane momentum. To also consider external screening effects resulting from, e.g., dielectric substrates or capping layers, coating molecules or other layered materials in the environment, we introduce the fully screened Coulomb interaction as:
	\begin{align}
		W_\text{env}^\text{TMDC}(\vec{q}, \omega) 
			= \frac{v(\vec{q})}{ \varepsilon_\text{env}^\text{TMDC}(\vec{q}, \omega) }. 
	\end{align}
	Here, the full dielectric function $\varepsilon_\text{env}^\text{TMDC}(\vec{q}, \omega)$, which accounts for the screening from the TMDC layer and from the environment, is derived using the WFCE approach\cite{rosner_wannier_2015}: we obtain $\varepsilon_\text{env}^\text{TMDC}(\vec{q}, \omega)$ by augmenting $\varepsilon^\text{TMDC}(\vec{q}, \omega)$ with the environmental screening $\varepsilon_\text{env}(\vec{q}, \omega)$ in the leading long-wavelength screening channel (see appendix \ref{appendix:fitting_parameter} and Ref.~\onlinecite{rosner_wannier_2015} for details).
	
	In order to study how environmental screening influences the band structure of the TMDC monolayer, we make use of the so-called $G\Delta W$ \cite{rohlfing_electronic_2010,winther_band_2017} approach. To this end, we start with an ab-initio $G_0W_0$ calculation for the free-standing monolayer utilizing the full band-structure (including a significant amount of unoccupied states) with $W_0$ corresponding to $W^\text{TMDC}$ as defined in Eq.~\eqref{eq:coulomb_FS}. 
	The resulting band structure is thus already affected by the internal screening processes of the TMDC layer itself. 
	We subsequently down-fold this $G_0W_0$ band structure to a minimal three-band / three-orbital model using adequately chosen localized Wannier functions. The resulting quasi-particle Hamiltonian and corresponding Green's function are called $H^\text{TMDC}$ and $G^\text{TMDC}$ in the following.
	The additional external screening effects described by $\varepsilon_\text{env}(\vec{q}, \omega)$ are subsequently added via
	\begin{align}
		\left[G^\text{TMDC}_\text{env}(\omega)\right]^{-1} = 
			\left[G^\text{TMDC}(\omega)\right]^{-1} + 
			\Sigma_{G\Delta W}(\omega), \label{eqn:Dyson}
	\end{align}
	with $G^\text{TMDC}(\omega) = [\omega \mathbb{1} - H^\text{TMDC} ]^{-1}$ and using the self-energy $\Sigma_{G\Delta W} = i G_0 \Delta W$ defined by the product of the non-interacting Green function $G_0$ (corresponding to the Kohn-Sham DFT results as used in the initial $G_0W_0$ step) and
	\begin{align}
		\Delta W(\vec{q}, \omega) 
			&= W_\text{env}^\text{TMDC}(\vec{q}, \omega) - W^\text{TMDC}(\vec{q}, \omega), \label{eqn:DW}
	\end{align}
	which is the difference between the full Coulomb interaction (including internal and external screening) and the Coulomb interaction of the free-standing TMDC. Similar to the standard $GW$ self-energy\cite{Hedin, hybertsen_electron_1986,Aryasetiawan_1998}, in the orbital (Wannier) basis $\Sigma_{G\Delta W}$ is defined by
	\begin{align}
		\Sigma_{G\Delta W}^{\alpha \beta} (\vec{k},\omega) =
		\sum_\lambda  
		\int d^2q \int \frac{d\omega'}{2\pi} 
		c_\alpha^{\lambda} (\vec{k}-\vec{q}) 
		[c_\beta^{\lambda} (\vec{k}-\vec{q})]^* \notag\\ 
		\times 
		\frac{2 \text{Im} \left[ \Delta W^{\alpha\beta} (\vec{q},\omega') \right]
			  \left[n_B (\omega^\prime) + n_F^{\lambda}(\vec{k}-\vec{q})\right]}
		{\omega + \omega' + i \delta - E_{\vec{k}-\vec{q}}^\lambda}, \label{eq:self_energy}
	\end{align}
	where $\alpha/\beta$ and $\lambda$ are orbital and band indices, respectively, $\delta$ is a small broadening, while $E_{\vec{k}}^\lambda$  and $c_\alpha^\lambda(\vec{k})$ are eigenenergies and expansion coefficients of the eigenfunctions of the $G_0W_0$ Hamiltonian $H^\text{TMDC}$, and $n_F^{\lambda}(\vec{k})$ and $n_B (\omega)$ are fermionic and bosonic occupation functions. Note, that the $G\Delta W$ self-energy defined in the equation above describes the \emph{changes} induced by the dielectric environment and does not explicitly involve the bare, purely real, and non-retarded Coulomb interaction $v_q$\cite{winther_band_2017}. The latter is unaffected by the environment and is already implicitly acounted for in $G^\text{TMDC}(\omega)$. We use Eq.~\eqref{eq:self_energy} to simulate situations with general retarded environmental screening.
	

	\begin{figure*}[htb]
		\includegraphics[width=\textwidth]{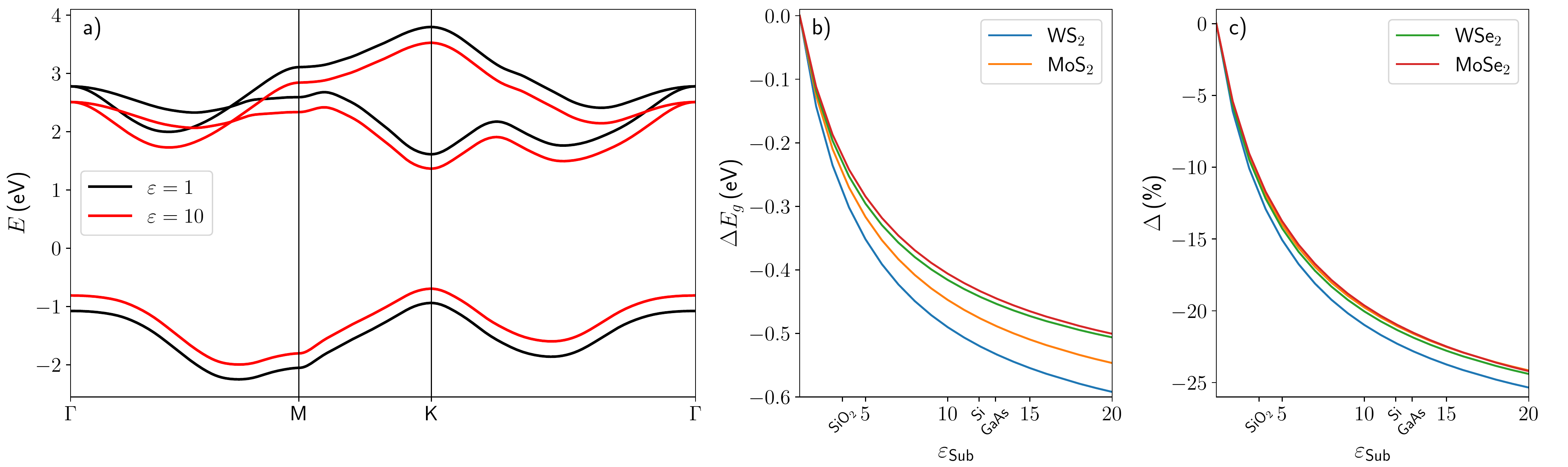}
		\caption{TMDC band structure modulations from substrates with instantaneous dielectric functions. (a) $G \Delta W$ band structure for WS$_2$ with $\varepsilon_{\t{sub}} = 1$ and $\varepsilon_{\t{sub}} = 10$ (without spin-orbit coupling). (b) Absolute and (c) relative differences of $G \Delta W$ band gaps (with spin-orbit coupling) compared to the band gap of freestanding TMDCs as functions of the dielectric constant $\varepsilon_{\t{sub}}$. \label{fig-2} }
	\end{figure*}
	
	
	The TMDC internal dielectric function is only weakly frequency dependent as long as $\omega$ is small compared to the TMDC band gap. In the case of instantaneous external screening $\varepsilon_\text{env}(\vec{q}, \omega) \approx \varepsilon_\text{env}(\vec{q})$ we can thus consider the total dielectric function 
	$\varepsilon_\text{env}^\text{TMDC}(\vec{q})$ to be frequency independent as well. This allows us to use the static Coulomb-hole plus screened-exchange approximation for the self-energy \cite{Hedin, hybertsen_electron_1986,Aryasetiawan_1998, PhysRevB.91.155109}  which reads in the orbital basis: 
	\begin{align}
		&\Sigma_{G\Delta W}^{\alpha \beta} (\vec{k}) = \label{eq:COHSEX} \\ \notag 
		&- \int \frac{d^2 q}{(2 \pi)^2} \sum_\lambda^{\t{occ}} \Delta W^{\alpha\beta} (\vec{q},\omega = 0) c_\alpha^\lambda (\vec{k}-\vec{q}) [c_\beta^\lambda (\vec{k}-\vec{q})]^* \\ \notag
		&+ \frac{1}{2} \int \frac{d^2 q}{(2 \pi)^2} \sum_\lambda \Delta W^{\alpha\beta} (\vec{q},\omega = 0) c_\alpha^\lambda (\vec{k}-\vec{q}) [c_\beta^\lambda (\vec{k}-\vec{q})]^*,	
	\end{align}
	where $c_\alpha^\lambda (\vec{k})$ are the coefficients of the DFT Hamiltonian in the orbital basis. The first so-called screened-exchange term  affects occupied states only and shifts all valence band states in energy. The second so-called Coulomb-hole term affects both, valence and conduction bands\cite{PhysRevB.91.155109, PhysRevLett.97.216405, PhysRevLett.102.046802, Waldecker2019}.
	As a result of the interplay between these two terms, the band-gap of the monolayer is reduced for negative $\Delta W(\vec{q}, \omega=0)$ and enhanced for positive $\Delta W(\vec{q}, \omega=0)$.
	Since the environmental screening always decreases $W_\text{env}^\text{TMDC}(\vec{q}, \omega=0)$ in comparison to $W^\text{TMDC}(\vec{q}, \omega=0)$, $\Delta W(\vec{q}, \omega=0)$ is always negative, so that any surrounding material will reduce the band gap as long as the static approximation holds.
	
	In the static approximation, Eq.~(\ref{eqn:Dyson}) allows us to define an effective Hamiltonian of the monolayer including the substrate screening effects according to $H_\text{env}^\text{TMDC}(\vec{k}) = H^\text{TMDC}(\vec{k}) + \Sigma_{G\Delta W}(\vec{k})$. The diagonalization of this Hamiltonian correspondingly yields the band structure of the monolayer as function of the environmental screening $\varepsilon_\text{env}$. 
	
	Spin-orbit coupling (SOC) can additionally be considered by a Russel-Saunders coupling with a k-dependent coupling parameter as described in Ref.~[\onlinecite{Steinhoff2014}]. The coupling parameters are chosen such that the SOC splittings at the valence- and conduction bands at high symmetry points match results of GGA calculations. 
	
	We get $W^\text{TMDC}(\vec{q}, \omega=0)$ in the orbital basis by projecting $W_0$ from our full $G_0W_0$ calculations in the Kohn-Sham basis to three Wannier orbitals, which are also used to represent $H^\text{TMDC}$. Afterwards, we calculate $W_\text{env}^\text{TMDC}$ using our WFCE approach\cite{rosner_wannier_2015}. Within the latter the additional environmental screening can expressed by simple dielectric constants $\varepsilon_\text{env}(\vec{q}, \omega) \equiv \varepsilon_\text{env}$ or full retarded dielectric functions $\varepsilon_\text{env}(\vec{q}, \omega)$ as resulting from substrates or capping layers (see Appendix \ref{appendix:fitting_parameter} for details).

\section{Results} \label{sec:results}
\subsection{Influence of instantaneous dielectric substrate screening} \label{sec:results_static}
	
	We start by investigating the situation depicted in Fig.~\ref{fig-1} (b), i.e. TMDC monolayers on homogeneous dielectric substrates. While the TMDC monolayers are described within the combined $G\Delta W$ and WFCE approaches on an ab initio level, the substrate screening is modeled in the following using simple but generic models which can be adjusted to realistc substrate material properties.
	In a first step, we consider local and instantaneous dielectric functions by setting $\varepsilon_\text{env}(\vec{q}, \omega) = \varepsilon_\text{sub}$, which is appropriate for bulk semiconducting substrates. This changes the long-wavelength limit of the total dielectric function to $\varepsilon_\text{env}^\text{TMDC}(\vec{q} \rightarrow 0) = (1 + \varepsilon_\text{sub})/2$ (see Ref. \onlinecite{PhysRevLett.102.076803}), while the short-wavelength behavior is unaffected $\varepsilon_\text{env}^\text{TMDC}(\vec{q} \rightarrow \infty) = \varepsilon^\text{TMDC}(\vec{q})$ (see Appendix \ref{appendix:fitting_parameter}) and allows us to use the static self-energy defined in Eq.~(\ref{eq:COHSEX}).
	
	In Fig.~\ref{fig-2}~(a) we show the band structure of free-standing ($\varepsilon_\t{sub}  = 1$) WS$_2$ together with the resulting band structure for $\varepsilon_\t{sub}  = 10$ without spin-orbit coupling in the minimal basis of the three transition d orbitals $d_{z^2}$, $d_{xy}$ and $d_{x^2-y^2}$. Upon increasing the environmental screening we decrease the Coulomb interaction and thus decrease the band gap. In more detail, we find a constant reduction of the gap between valence and conduction bands throughout the whole Brillouin zone. This ``scissor-like'' behavior is a direct result of the non-local screening the TMCD monolayer is exposed to and which leads to a strongly peaked $\Delta W(\vec{q})$ in momentum space as discussed for the example of WS$_2$ in detail in Ref.~[\onlinecite{Waldecker2019}]. 

	In the following, we concentrate on the comparison between different TMDCs and their reactions to their dielectric environments. To this end we calculate the band gaps $E_g(\varepsilon_\text{sub})$ (considering spin orbit coupling) for different dielectric constants for all four TMDCs and show the absolute band-gap differences $\Delta E_g(\varepsilon_\text{sub}) = E_g(\varepsilon_\text{sub})-E_g(\varepsilon_\text{sub}=1)$ as well as the relative ones $\Delta(\varepsilon_\text{sub}) = \left( \frac{E_g(\varepsilon_\text{sub})-E_g(\varepsilon_\text{sub}=1)}{E_g(\varepsilon_\text{sub}=1)} \right)$ in Fig.~\ref{fig-2} (b) and Fig.~\ref{fig-2} (c), respectively. 
	
    We see significant band-gap reductions with increasing environmental screening for all TMDCs. For $\varepsilon_\t{sub} = 5$ the band gaps are reduced by about $300$ to $350\,$meV depending on the specific material [c.f. Fig.~\ref{fig-2} (b)]. Realistic substrates, such as SiO$_2$ or Si have macroscopic dielectric constants of about $3.6$ [\onlinecite{Robertson2004_high_dielectric}] and $12$ [\onlinecite{Sze2006_constants}], respectively, yielding reductions of up to $500\,$meV. 
	
	The sulfides are slightly stronger affected than the selenides with larger absolute changes in their band gaps, as was also found by Winther and Thygesen for the comparison between MoS$_2$ and MoSe$_2$ \cite{Winther2017}. Compared to the MS$_2$, the MSe$_2$ compounds have smaller band gaps and thus exhibit larger \textit{internal} polarizabilities so that changes in the external screening affect the total screening in the selenides less than in the sulfides. However, these differences are of quantitative rather than qualitative nature. Indeed, the relative substrate-induced band-gap reductions as shown in Fig.~\ref{fig-2}~(c) are very similar for all materials and amount to about $15$\% for $\varepsilon_\t{sub} = 5$.

\subsection{Frequency-dependent substrate screening} \label{sec:results_dynamic}

	\begin{figure}
		\includegraphics[width=0.99\columnwidth]{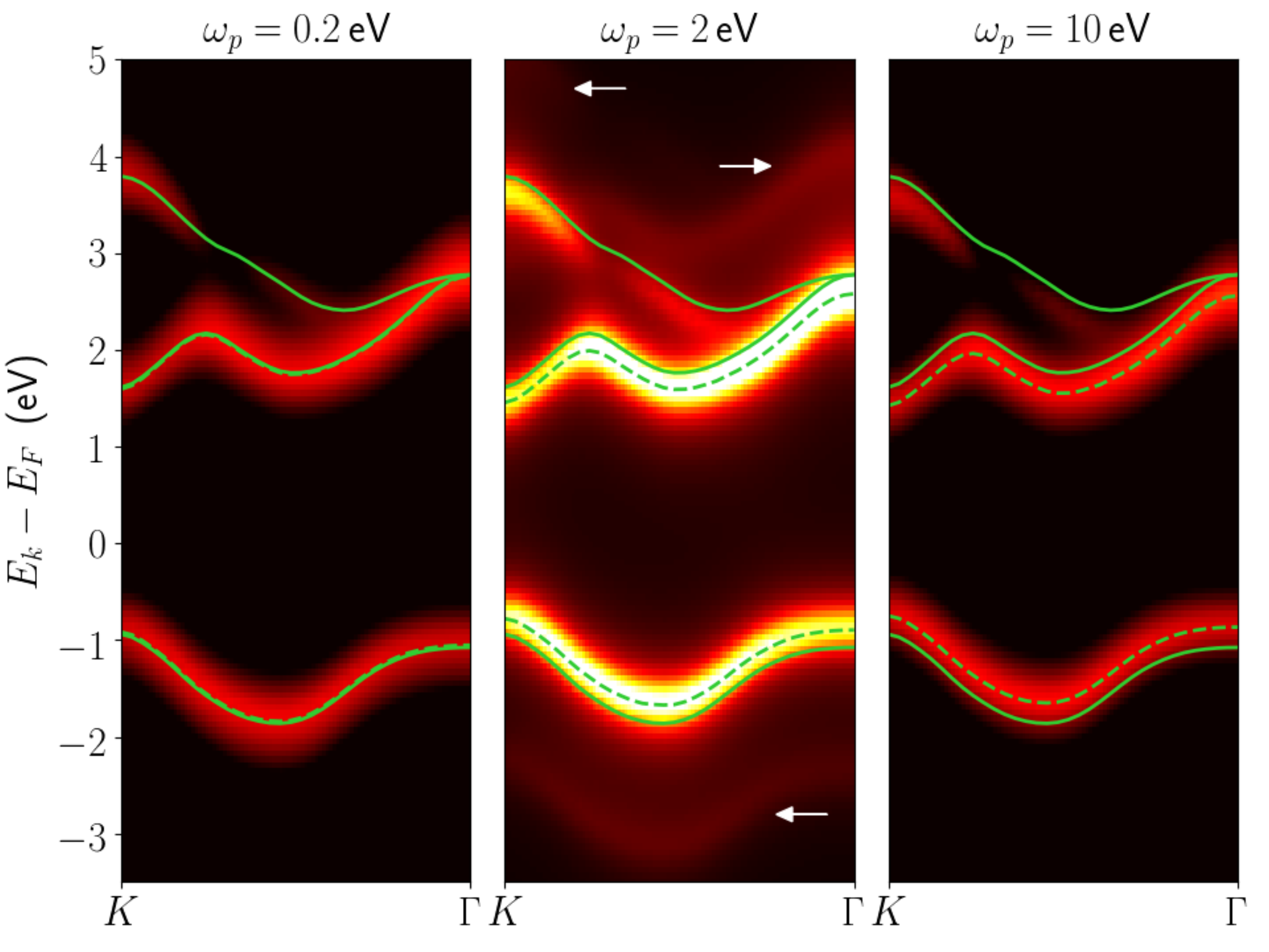}
		\caption{WS$_2$ spectral functions for different substrate resonance frequencies $\omega_p$ and fixed $\varepsilon_\text{sub}^{(0)}= 10$ together with the free-standing band structure (solid lines). The dashed lines show the substrate-screened quasi-particle band structure resulting from Gaussian fits to the spectral function. Bosonic side bands resulting from coupling between the TMDC electrons and substrate excitations are marked by arrows.
		} \label{fig-3}
	\end{figure}
	
	\begin{figure}
		\includegraphics[width=\columnwidth]{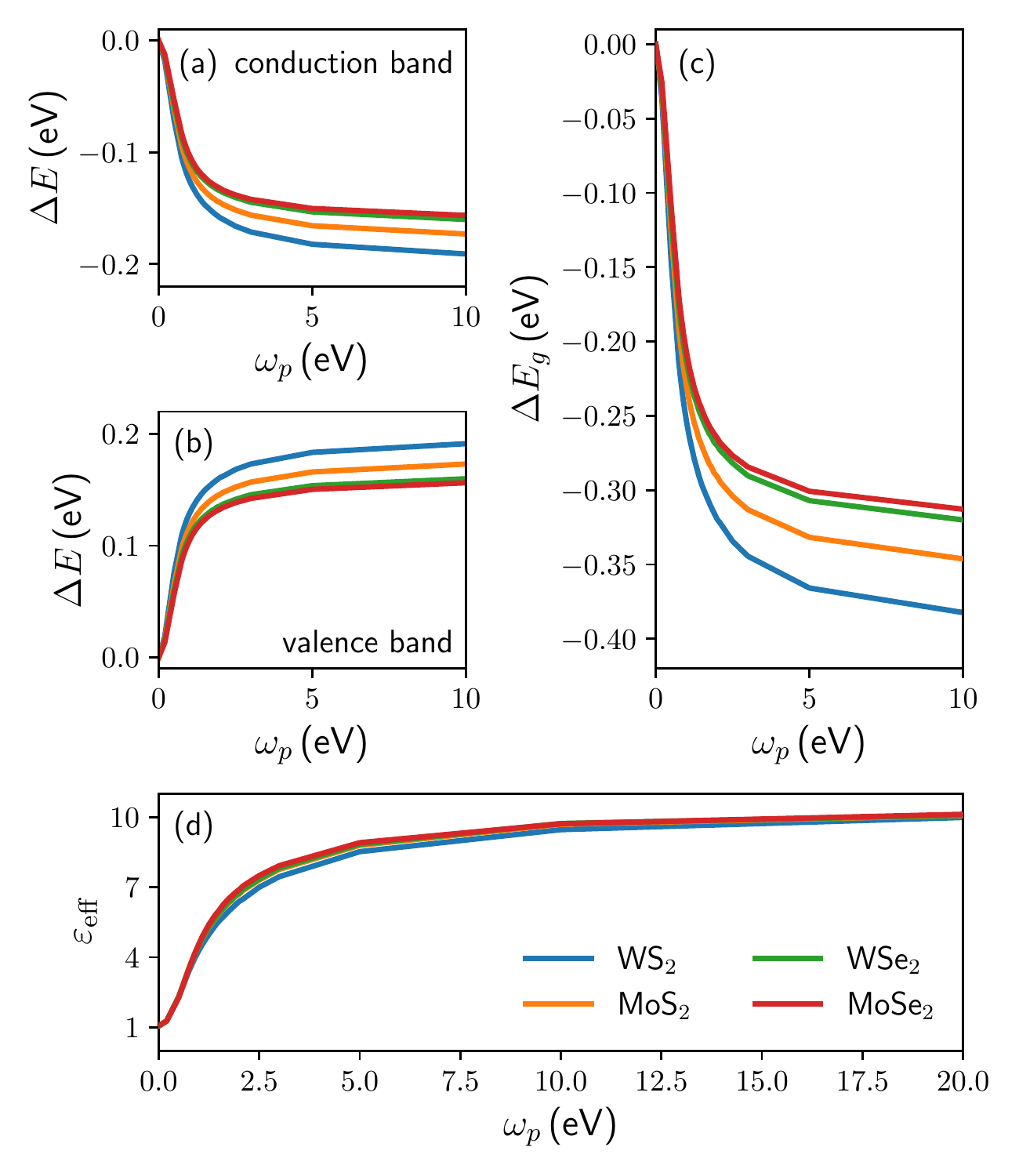}
		\caption{Substrate resonance frequency dependent (a) valence and (b) conduction band renormalizations at $\vec{K}$ for all TMDCs under investigation. (c) Total band-gap modifications $\Delta E_g (\omega_p)$ and (d) effective static dielectric functions $\varepsilon_\t{eff} (\omega_p)$.} \label{fig-4}
	\end{figure}

	In realistic experimental situations the screening by substrates, capping layer, or molecular adsorbates will be dynamic, i.e. the external dielectric function will be frequency dependent. Generally, phonons\cite{AlexSteinhoff2019} and (interband) plasmons \cite{Steinhoff2018, PhysRevB.34.979, PhysRevB.55.13961} contribute to this frequency dependence. While our formalism is general, we focus in the following on the effects of interband plasmons in the materials surrounding the 2D TMDC layer. 
	To this end, we make use of a plasmon-pole model \cite{Engel1993,Hwang2018} of the form
    \begin{align}
     \frac{1}{\varepsilon_\t{sub} (\omega)} = 1 + \frac{A}{\pi \braq{\brak{\omega + i \eta}^2 - \omega_p^2}}, \label{eq:ppm}
    \end{align}
	with a small broadening $\eta = 0.1\,$eV and $A = \pi \omega_p^2 [ 1 - 1/\varepsilon^{(0)}_\t{sub}]$, where we introduce $\omega_p$ as a \textit{single} substrate resonance frequency and the static limit of the substrate dielectric function $\varepsilon_\t{sub} (\omega=0) = \varepsilon^{(0)}_\t{sub}$ as model parameters.
	Large $\omega_p$ as compared to all electronic TMDC energies, and in particularly compared to the TMDC band gap and band width, leads to the anti-adiabatic limit $\varepsilon_\t{sub} (\omega) \approx \varepsilon^{(0)}_\t{sub} $ at all $\omega$ of interest, which is covered by the static approximation discussed before. In the limit of small $\omega_p$ we regain the freestanding monolayer situation, i.e. ${\lim_{\omega_p \rightarrow 0} \varepsilon_\t{sub} (\omega) = 1}$ at finite $\omega > 0$. Generally, in between, we need to utilize the full self-energy defined in Eq.~(\ref{eq:self_energy}).

	In Fig.~\ref{fig-3} we show the resulting spectral functions for WS$_2$ for different substrate resonance frequencies $\omega_p$ and fixed $\varepsilon^{(0)}_\t{sub} = 10$ together with the free-standing band structure. We see the three cases: for small $\omega_p = 0.2\,$eV the conduction and valence bands are nearly unaffected, as the substrate-screened quasi-particle band structure (dashed lines) falls on top of the freestanding one (solid lines).
	For large $\omega_p = 10\,$eV we find the strongest, symmetric renormalization as discussed in the previous parts. 
	For intermediate $\omega_p = 2$\,eV, which is on the order of the monolayer band gap, we see smaller renormalizations and side bands resulting from the coupling between the TMDC electrons and bosonic substrate excitations. For the valence bands these accompanying satellite bands appear at lower frequencies (shifted by about $-\omega_p$), and for the conduction bands at higher frequencies (shifted by about $+\omega_p$) \cite{Giustino2017}. 

	In Fig.~\ref{fig-4} we show the renormalization of the valence (v) and conduction (c) band edges
	\begin{align}
		\Delta^\text{v/c} (\omega_p)= E_\vec{K}^{\text{FS},\text{v/c}} - E_\vec{K}^{\varepsilon,\text{v/c}} (\omega_p)
	\end{align}
	and the total band-gap modification $\Delta E_g (\omega_p) = \Delta^\text{c}(\omega_p) -\Delta^\text{v}(\omega_p)$ in dependence of $\omega_p$, where $E_\vec{K}^{\text{FS},\text{v/c}} (\omega_p)$ and $E_\vec{K}^{\varepsilon,\text{v/c}} (\omega_p)$ are the renormalized quasi-particle energies at $\vec{K}$ of the freestanding monolayer and for the substrate-screened one, respectively. We see a negative shift of the conduction band and a positive shift in the valence band, yielding a decreasing band gap with increasing $\omega_p$. The band gap is always symmetrically reduced and the screening induced changes are generally slightly bigger in the sulfides than in the selenides, as discussed for the static dielectric function above. 

	\begin{figure*}[tbh]
		\includegraphics[width=0.33\textwidth]{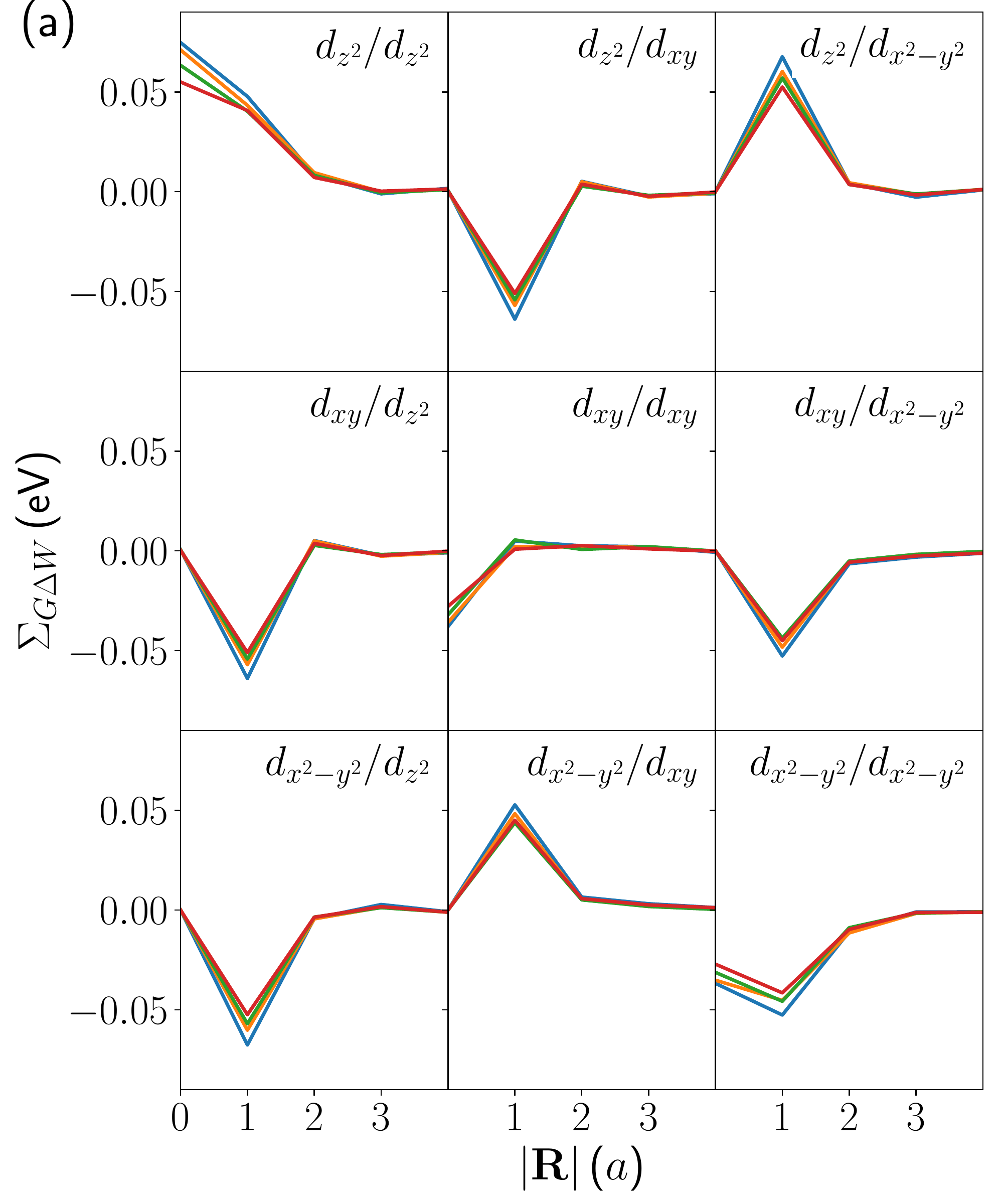}\hfill
		\includegraphics[width=0.33\textwidth]{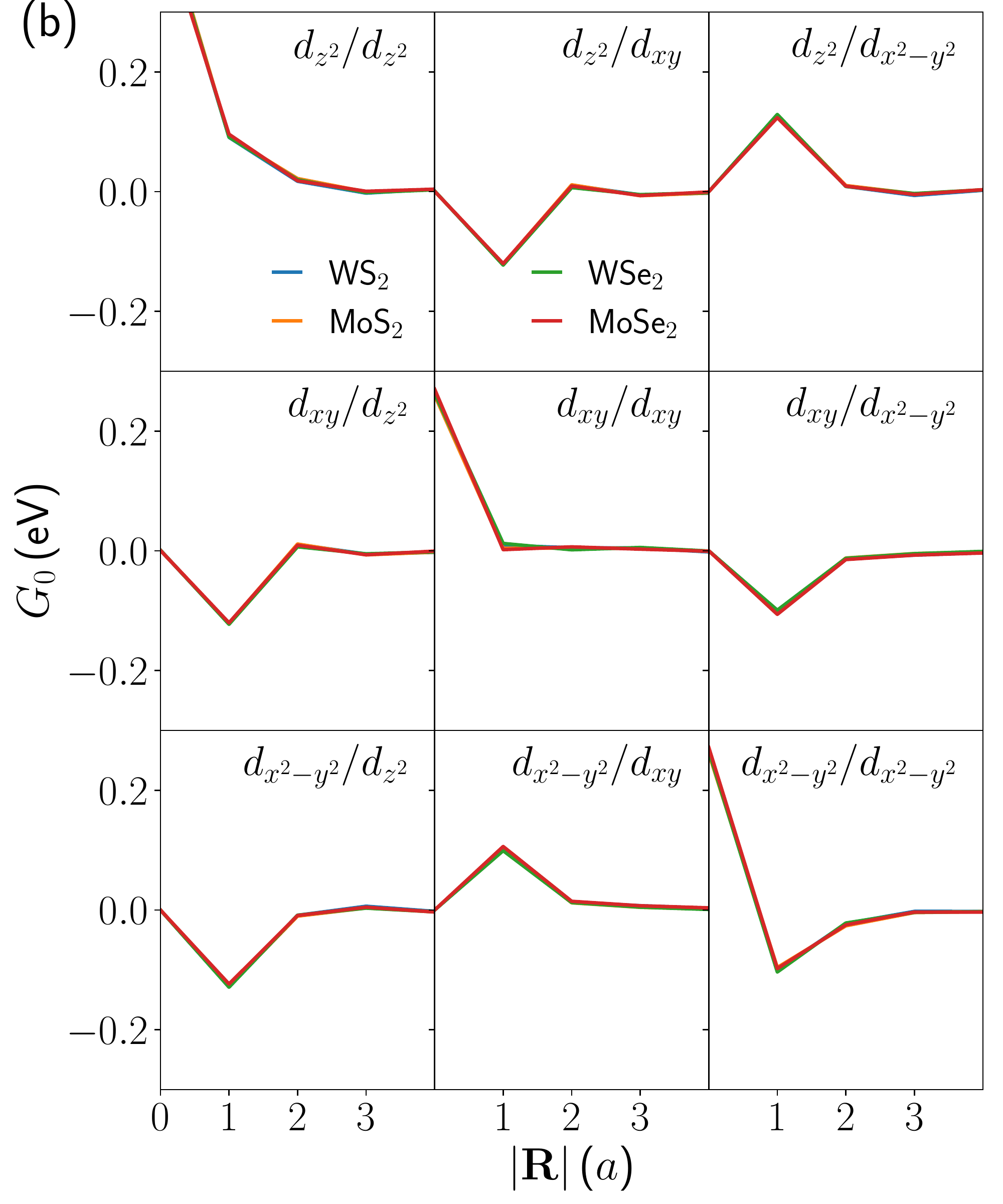}\hfill 
		\includegraphics[width=0.33\textwidth]{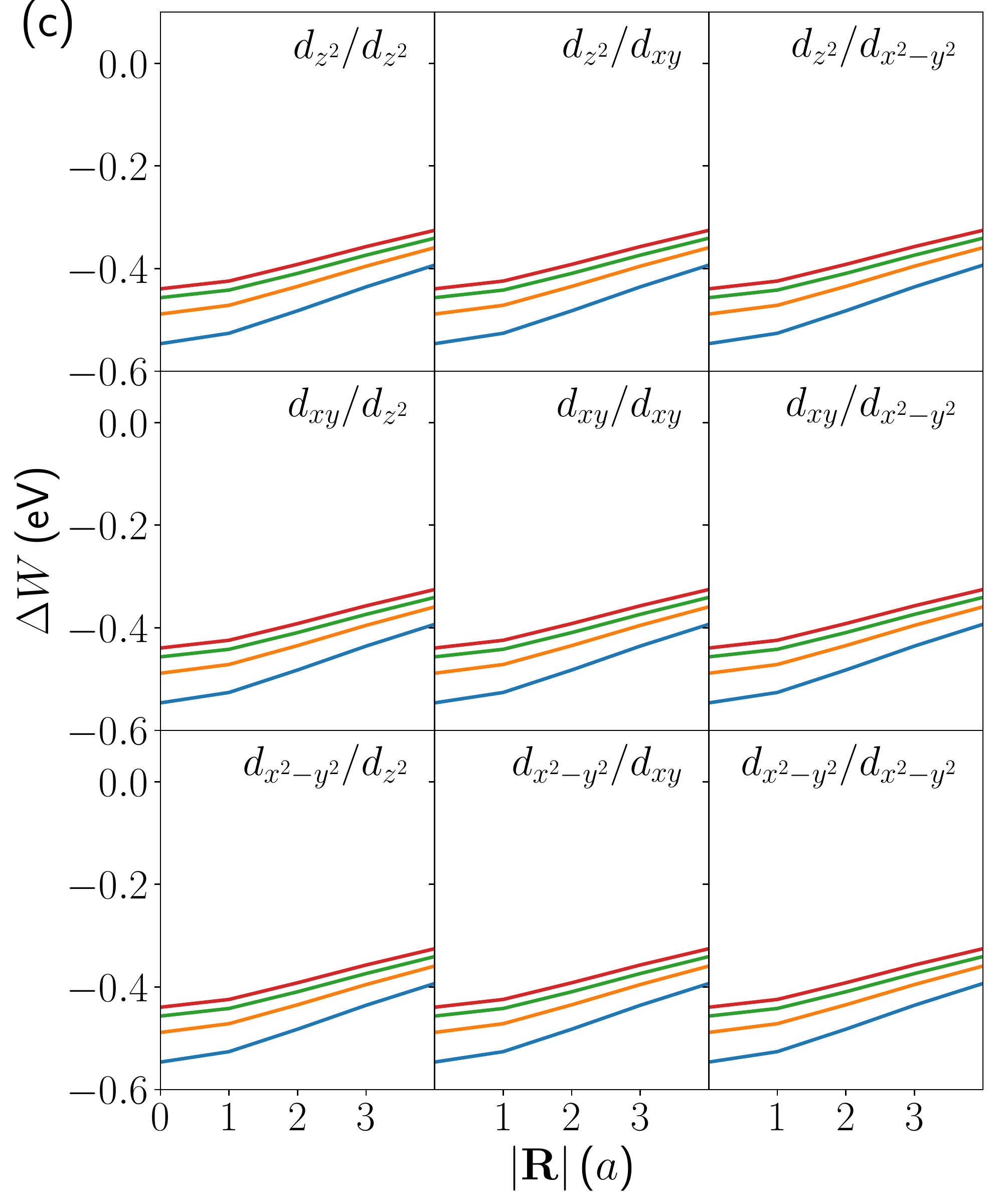}
		\caption{(a) $\Sigma^{\alpha\beta}_{G\Delta  W}(\vec{R},\omega=0)$, (b) $G_0^{\alpha\beta}{G\Delta  W}(\vec{R},\omega=0)$, and (c) $\Delta W^{\alpha\beta}{G\Delta  W}(\vec{R},\omega=0)$ along the real-space direction $\vec{a}_1 = [a,0]$ in units of the lattice constant $a$ for all investigated TMDCs on a substrate with an effective dielectric constant $\varepsilon_\text{env} = 10$.}  \label{fig:self_energy} 
	\end{figure*}

	Based on these monotonous symmetric shifts we can define for each frequency $\omega_p$ an effective \textit{static} dielectric constant $\varepsilon_\t{eff}(\omega_p)$ which leads to the same renormalization of the quasi-particle band structure as the frequency-dependent external dielectric function. To this end, we combine the data from Fig.~\ref{fig-4}~(c) and Fig.~\ref{fig-2}~(b). The resulting effective $\varepsilon_\t{eff}(\omega_p)$ is shown in Fig.~\ref{fig-4}~(d). Both limits of the plasmon-pole model can be clearly seen: in the lower limit of $\omega_p$ the substrate screening vanishes $\varepsilon_\t{eff}(\omega_p \rightarrow 0) \rightarrow 1$ so that the band structures are not affected, as seen in Fig.~\ref{fig-3}~(a). For large $\omega_p$, $\varepsilon_\text{eff}$ recovers the static dielectric constant $\varepsilon_\t{eff}(\omega_p) = \varepsilon^{(0)}_\t{sub} = 10$.  
	Noticeable, the effective dielectric constant $\varepsilon_\t{eff}(\omega_p)$ has nearly the same $\omega_p$-dependence for all four TMDCs and thus does not show any strong material dependencies.
	
	A mapping of this kind can generally be performed  as long as the quasi-particle approximation holds, idependently of the number of poles in $\varepsilon_\text{sub}(\omega)$. Multiple poles in $\varepsilon_\text{sub}(\omega)$ will, however, likely create a more complicated dependence of $\varepsilon_\t{eff}$ on the pole positions and strengths. Also, it is important to note that this mapping is optimized to reproduce the single-particle band gap and might not be appropriate to capture modifications to two-particle excitations, such as excitons, induced by the dynamic substrate screening.
	
	For our further discussions of the Coulomb engineered heterostructures and their spatially varying band gaps we can thus stick to the static limit of the $G\Delta W$ approach utilizing effective instantaneous dielectric constants $\varepsilon_\t{eff}$ as along as we assume that the relevant substrate plasmon frequencies entering $\varepsilon_\text{sub}$ are in the optical frequency range. 

\subsection{Self-energy length scales} \label{sec:results_length}

	For Coulomb engineered heterostructures not only the band gap reduction but also the \textit{length scale} on which this reduction takes place is important. As the extent of the self-energy is an intrinsic measure for how sharp an interface in a Coulomb engineered heterostructure as depicted in Fig.~\ref{fig-1}~(c) can be, we discuss this length scale for TMDC monolayers in the following. 
	
	In Fig.~\ref{fig:self_energy}~(a) we show the self energies for all four semiconducting TMDCs and using $\varepsilon_\t{env} = 10$ as the substrate dielecrtric constant. The self energies are plotted along the real-space direction $\vec{a}_1 = [a, 0]$ with $a$ being the lattice constant and for all orbital channels.

	In the static Coulomb-hole plus screened-exchange approximation $\Sigma_{G\Delta W}$ can also be interpreted as the \textit{renormalization} of the hopping matrix elements of the TMDC Hamiltonian due to screening effects from the environment. We find that all hopping elements are renormalized due to screening effects, i.e. intra- ($\Sigma_{G\Delta W}^{\alpha \alpha}$) as well as inter-orbital ($\Sigma_{G\Delta W}^{\alpha \beta}$) terms. 
	Most importantly we find not only local renormalizations $\Sigma_{G\Delta W}^{\alpha\beta} (\vec{R} = 0)$ but especially non-local hopping terms are changed. The renormalization due to non-local inter-orbital terms change the hybridization of the system and are mainly responsible for the change of the band gap of TMDCs in dielectric environments (as was also discussed in c.f. Ref.~[\onlinecite{rosner_two-dimensional_2016}]). 
	In fact, we see that the local diagonal elements $\Sigma_{G\Delta W}^{\alpha\alpha} (\vec{R} = 0)$ have opposite signs (positive for $\alpha = d_{z^2}$ and negative for $\alpha = d_{xy} / d_{x^2-y^2}$), which shift $d_{z^2}$ states up in energies and $d_{xy} / d_{x^2-y^2}$ states down. The local terms alone (note that off-diagonal local, i.e. $\vec{R} = 0$, terms are zero) would thus \textit{enhance} the band gap at $\vec{K}$ upon \textit{increasing} the environmental screening\footnote{At $\vec{K}$ the upmost valence band is of $d_{xy} / d_{x^2-y^2}$ and the lowest conduction band of $d_{z^2}$ character}.
	To realistically describe the modifications of the band structure and the band gap it is thus important to capture non-local effects.
	The largest contributions for all orbital combinations and TMDCs can be found within two unit cells which corresponds to a distance of roughly 6.2\,\AA\ to 6.6\,\AA\ (depending on the material). This length scale is similar for all investigated materials. 
	
	In real space, the self-energy is the direct product of the Coulomb interaction $\Delta W(\vec{R}, \omega)$ and the non-interacting Green's function $G_0(\vec{R}, \omega)$ convoluted in the frequency domain\cite{Berger2012}
	\begin{align}
		\Sigma_{G\Delta W}(\vec{R}, \omega) \notag =
		   i \int_{-\infty}^{+\infty} \frac{d\omega'}{2 \pi} \, e^{i \eta \omega'}
		   G_0(\vec{R}, \omega + \omega') \Delta W(\vec{R}, \omega'). \notag
	\end{align}
	As $\Delta W$ is a strongly peaked function in momentum space around $\vec{q} = 0$, it is nearly constant in real space as can be seen in Fig.~\ref{fig:self_energy}~(c). Hence, the spatial decay of the self-energy must result from properties of the Green's function.
	In Fig.~\ref{fig:self_energy}~(b) we show $G_0^{\alpha \beta} (\vec{R},\omega=0)$ . The overall curve characteristic resembles the self-energy and shows the same length scale. Thus the spatial extent of the self-energy is determined by the spatial extent of $G_0$, which turns out to be the decisive material specific property to determine the intrinsic length scale of a Coulomb engineered heterostructure. As the semiconducting TMDCs under consideration have similar electronic band structures, they consequently have similar extents of $G_0$ which explains the similar length scales of the self-energies. Thus, Coulomb engineered heterostructures should allow for spatial band gap variations within a few lattice constants\cite{rosner_two-dimensional_2016} in all semiconducting TMDCs.

\subsection{Coulomb engineered heterostructures} \label{sec:results_heterostruct}

	\begin{figure}[htb]
	\begin{minipage}{0.5\textwidth}
		\includegraphics[width=0.95\textwidth]{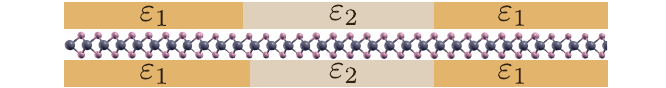}\\
		\includegraphics[width=\columnwidth]{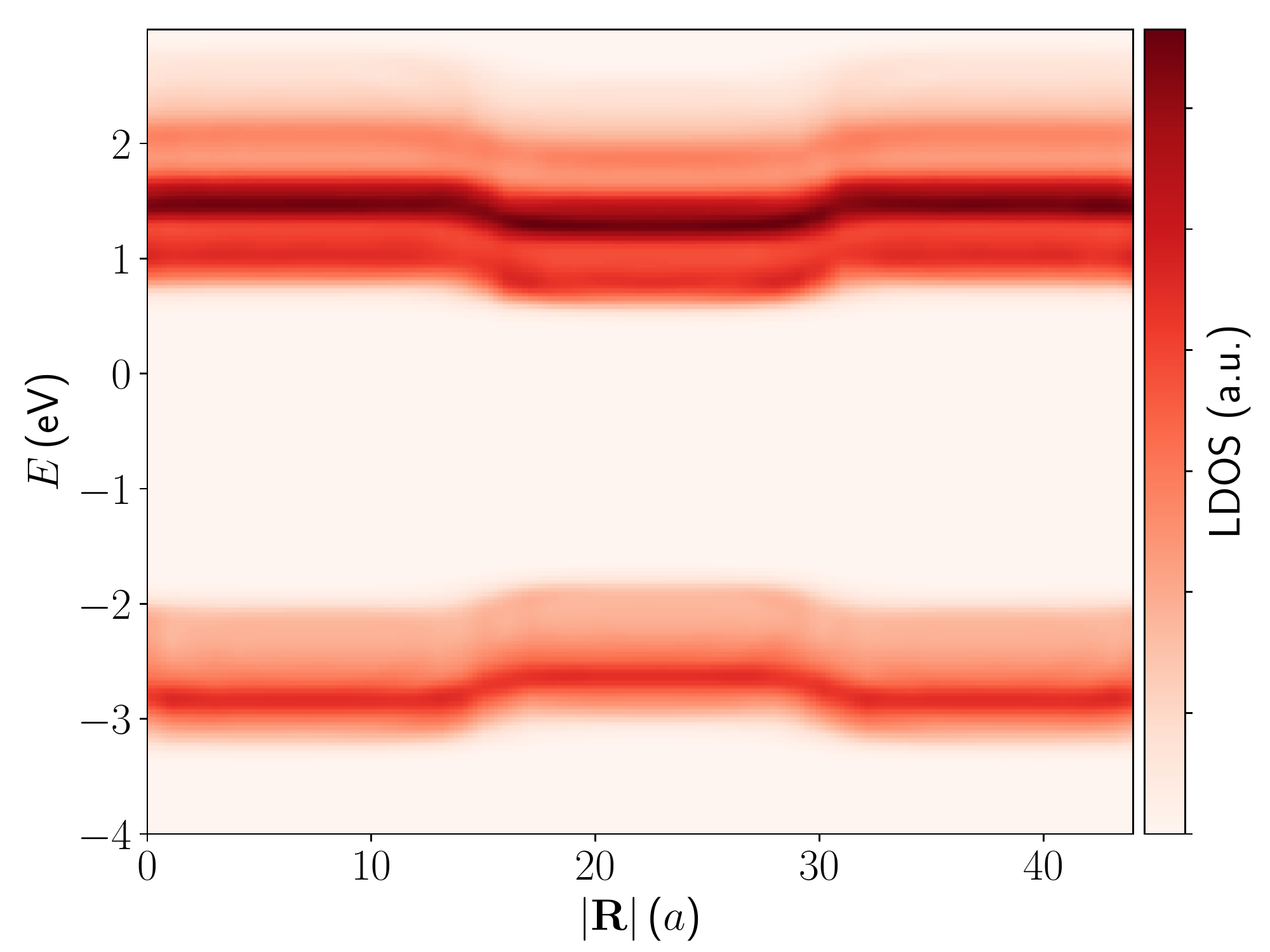}
	\end{minipage} \hfill
	\begin{minipage}{0.5\textwidth}
		\caption{Top: Model for Coulomb-engineered heterostructure using spatially structured screening environments with dielectric constants $\varepsilon_1$ and $\varepsilon_2$. Bottom: Local density of states for a Coulomb-engineered heterostructure using WS$_2$ with $\varepsilon_1 = 1$, $\varepsilon_2 = 10$ and $\varepsilon_3 = 1$. \label{fig-6} }
	\end{minipage}
	\end{figure} 

	In the following we aim to describe Coulomb engineered heterostructures from spatially structured dielectric environments as, for example, shown in Fig.~\ref{fig-1}~(c). Describing such systems is numerically very challenging due to the broken translational symmetry perpendicular to the interface of the substrate. However, as we showed above, the self-energy $\Sigma_{G\Delta W}$ is short ranged and we can use a static dielectric constant $\varepsilon_\t{eff}$ for the description of the substrate screening effects to the hopping matrix elements of the TMDC. We thus model the self-energy of a Coulomb engineered heterostructure with the help of the self-energies from homogeneous calculations:
	\begin{align}
			\Sigma_{\alpha \beta}^\text{het} (\vec{R} ) = 
				\begin{cases}
	              \Sigma_{\alpha \beta}^{\varepsilon_1} & \text{hopping in } \varepsilon_1 \\
	              \Sigma_{\alpha \beta}^{\varepsilon_2} & \text{hopping in } \varepsilon_2 \\
	              \frac{1}{2} (\Sigma_{\alpha \beta}^{\varepsilon_1} + \Sigma_{\alpha \beta}^{\varepsilon_2} ) & \text{hopping between } \varepsilon_i
	            \end{cases}
	\end{align}
	mimicking an abrupt change in the dielectric environment. Here $\Sigma_{\alpha \beta}^{\varepsilon_i}$ is calculated from Eq.~\eqref{eq:COHSEX} and Fourier transformed to real space.
	With that we get the Hamiltonian for the full heterostructure:
	\begin{align}
		H_{\alpha \beta}^\text{het}(\vec{R} ) 
			= H_{\alpha \beta}^{\text{FS}} (\vec{R} ) 
			+ \Sigma_{\alpha \beta}^\text{het}(\vec{R} )
	\end{align}
	using the free-standing Hamiltonian $H_{\alpha \beta}^{\text{FS}}(\vec{R})$ from our initial $G_0W_0$ calculation.
	The resulting local density of states (LDOS) for WS$_2$ encapsulated by a sub/superstrate with two dielectric interfaces ($\varepsilon_{1} = 1$ and $\varepsilon_2 = 10$) is shown in Fig.~\ref{fig-6}. Darker red areas depict high density of states whereas the light red area corresponds to nearly zero LDOS indicating the gapped region around $E=0$. The band gap modulation in the different areas is clearly visible and we see a symmetric band alignment, as already described in Fig.~\ref{fig-2}~(a). The change of the band gap from one region to the other is limited to a few unit cells as expected from the spatial extent of the self-energy. These results are different to our previous model calculation presented in Ref.~[\onlinecite{rosner_two-dimensional_2016}], where we used the Hartree-Fock approximation which effectively neglects the Coulomb-hole part of the Coulomb-hole plus screened-exchange self-energy used here. Thus, our previous calculations showed changes to the valence band only. Taking the full self-energy into account we hence find a spatial band gap modulation reminiscent of type-I heterojunctions. 
	It is worth noting, that next to the spatial extent of the self-energies also image-charge effects might affect the length scale on which the band gap is modulated in these heterostructures. While we included these effects directly in our previous Hartree-Fock calculations\cite{rosner_two-dimensional_2016}, we handle them here approximately by assuming piecewise constant real-space self-energy terms on either side of the interface. However, as we discuss in Appendix \ref{appendix:HS_IC} these image-charge induced modulations of the band gap variation length scale will not qualitatively change the conclusions drawn here. Finally, we would like to highlight that in the depicted case with two dielectric interfaces we can even imprint a quantum-wire-like structure to the active TMDC layer.

\section{Conclusions}

	Based on a combination of the $G\Delta W$ and WFCE approaches we were able to develop a material-realistic description of Coulomb-engineered heterojunctions in semiconducting TMDC monolayers. We found that all investigated TMDCs are similarly susceptible to screening-induced band-gap reductions, which can be on the order of several hundred meV. Retardation effects in the environmental screening as expressed by the frequency dependence affect the magnitude of the band gap renormalizations in such a way, that dielectric environments with high plasmon frequencies turn out to be most effective for external band structure manipulations. The electronic quasi-particle band structures in presence of frequency-dependent external dielectrics can be described in terms of effective instantaneous dielectric functions. From an analysis of the self-energy in real-space we showed, that the spatial extent of the self-energy is a material-intrinsic property. In the case of semiconducting TMDCs this spatial extent is limited to neighboring unit cells. This localization together with the effective handling of retarded environmental screening effects, allowed us to derive a tight-binding based modeling scheme to describe Coulomb-engineered heterojunctions resulting from dielectric interfaces in the substrate. Based on these material-realistic simulations we found that spatial band-gap modulations reminiscent of type-I heterojunctions can be externally and non-invasively induced in a monolayer of WS$_2$. This renders TMDCs promising candidates for future applications based on Coulomb engineering. 

\begin{acknowledgments}
    This work was supported by the DFG via RTG 2247 and \mbox{CZ 31/20-1} and by computing resources at HLRN. 
	We acknowledge fruitful discussions with Andrew J. Millis and Hugo Strand and thank Merzuk Kaltak for sharing his (c)RPA implementation\cite{Kaltak2015}. MR thanks the Alexander-von-Humboldt Foundation for support. 
\end{acknowledgments}

\appendix

\section{Calculation details for homogeneous monolayer TMDCs} \label{appendix:Ai_calculation}

All ab-initio calculations were performed within the Vienna ab initio simulation package (\texttt{VASP}) \cite{Kresse_Hafner_1993, Kresse1996}. The DFT calculations were carried out within the GGA approximation \cite{perdew_generalized_1996} utilizing a PBE plane wave basis set with a cutoff energy of 350\,eV. All structures were relaxed on $18\times18\times1$ k-meshes until the total free energy change was smaller than $10^{-4}$\,eV. The resulting parameters are listed in Tab.~\ref{table:fit_parms}. $G_0W_0$ calculations were performed on $24 \times 24 \times 1$ k-meshes, using $280$ bands, $\omega$ grids with $200$ grid points within \texttt{VASP}'s default $\omega$ limits, $G_0W_0$  cutoff energies of $150\,$eV, and interlayer distances of $20\,$\AA\ yielding a good compromise between numerical feasibility and accuracy. Due to the applied super-cell approach, the $G_0W_0$ results are affected by artificial self-interactions between periodic images of the 2D layer, which yields underestimated band gaps (see Tab.~\ref{table:fit_parms}) in comparison to fully converged results \cite{Haastrup2018}. However, we focus on band gap changes only, which are essentially converged in the computational setup chosen here (see Appendix \ref{appendix:gdw_convergence}). 

The freestanding Hamiltonian $H^\text{TMDC}$ is described in a Wannier basis by projecting the $G_0W_0$ results to d$_z^2$, d$_{xy}$, and d$_{x^2-y^2}$ orbitals using the \texttt{Wannier90} \cite{mostofi_updated_2014} code. Appropriate inner energy windows are chosen to include the highest valence band and as much of the lowest conduction bands as possible. We do not perform any maximal localization and use only the disentanglement procedure to maintain the dominant orbital characters. 

The spin orbit coupling was considered afterwards by a Russel Saunders coupling \cite{Steinhoff2014} with a k-dependent coupling parameter 
\begin{align}
 \lambda(\vec{k}) = \lambda_0 \cdot e \cdot \brak{1 - \frac{\abs{\vec{k}-\vec{K}}}{\abs{\vec{K}}}}^2 \cdot e^{-\brak{1 - \frac{\abs{\vec{k}-\vec{K}}}{\abs{\vec{K}}}}^2}
\end{align}
which was chosen such that spin dependent GGA band structures are reproduced. The parameter $\lambda_0$ is equal to the spin splitting $\Delta_\t{SOC}$ at the $\vec{K}$ point of the GGA band structures and is listed in Tab.~\ref{table:fit_parms}.

\section{Modeling the Coulomb interaction: Method and Parameter} \label{appendix:fitting_parameter}

\begin{table*}[t]
\begin{tabular}{c||c|c|c|c|c|c|c|c|c|c|c}
         & $E_g^0$ & $a$  	& $z_0$ 	& d     				   & $\Delta_\text{SOC}$ & $\varepsilon_\infty$ & $\varepsilon_2$ & $\varepsilon_3$ & $\gamma$ & $V_2$ & $V_3$ \\
         & in eV   & in \AA 	& in \AA 	& in \AA				   & in eV    &                      &                 &                 & in \AA   & in eV & in eV \\ \hline \hline
MoS$_2$  & 2.26    & 3.18 	& 3.13    & 6.148 \cite{Coehoorn1987} & 0.148    & 10.136               & 2.637           & 2.019           & 1.990    & 0.817 & 0.360 \\
MoSe$_2$ & 2.07    & 3.32	&3.34	& 6.450 \cite{Coehoorn1987} & 0.186    & 11.282               & 2.307           & 1.787           & 1.637    & 0.867 & 0.402 \\
WS$_2$   & 2.33    & 3.19 	&3.15	& 6.162 \cite{Schutte1987}  & 0.427    & 8.565                & 2.913           & 2.281           & 2.169    & 0.737 & 0.332 \\
WSe$_2$  & 2.07    & 3.32	&3.36	& 6.480 \cite{Schutte1987}  & 0.464    & 9.873                & 3.097           & 2.490           & 2.733    & 0.647 & 0.303
\end{tabular}
\caption{Relaxed lattice constant $a$, distsance between chalcogen atoms $z_0$, interlayer distance $d$ and fit parameter of the bare Coulomb interaction ($\gamma$, $V_2$,$V_3$) as well as for the dielectric function ($\varepsilon_\infty$, $\varepsilon_2$, $\varepsilon_3$). Additionally, we show the spin-orbit coupling parameter determined from the valence band splitting at $\vec{K}$ in GGA calculations.}\label{table:fit_parms}
\end{table*}

We utilize our WFCE approach\cite{rosner_wannier_2015} to (A) analytically describe all involved Coulomb interaction matrix elements $W^{\alpha \beta}$ in the orbital basis and (B) to include the external dielectric screening effects.
To this end, we fit the density-density matrix elements of the bare Coulomb interaction $v^{\alpha \beta}(\vec{q})$ and the dielectric function $\varepsilon^{\alpha \beta}(\vec{q}, \omega=0)$ of the freestanding monolayers calculated from first principles in RPA (using a recent \texttt{VASP} implementation by Kaltak\cite{Kaltak2015}) to analytic functions as described in Ref.~[\onlinecite{schonhoff_interplay_2016}].

	We start with diagonalizing the bare interaction $v(\vec{q})$ in the Wannier orbital basis yielding 
	\begin{align}
            v^\t{diag}(\vec{q}) = \sum_{i=1}^3 v_i(\vec{q}) \ket{e_i}\bra{e_i},
	\end{align}
        where $v_1(\vec{q}) = \braket{e_1|v(\vec{q})|e_1}$ is the leading (i.e, largest) eigenvalue and $e_i$ are the eigenvectors of $v(\vec{q})$ in the long-wavelength limit $q \rightarrow 0$:
        \begin{align}
         e_1 = \begin{pmatrix}
                1/ \sqrt{3} \\ 1/ \sqrt{3} \\ 1/ \sqrt{3}
               \end{pmatrix}, \quad 
               e_2 = \begin{pmatrix}
                +4/\sqrt{6} \\ -1/ \sqrt{6} \\ -1/ \sqrt{6}
               \end{pmatrix}, \quad
               e_3 = \begin{pmatrix}
                0\\ +1/ \sqrt{2} \\ -1/ \sqrt{2}
               \end{pmatrix}.
        \end{align}
        The leading eigenvalue can be interpreted as long wavelength charge-density modulations to which screening effects due to environments are supposed to be strongest. For this limit, a macroscopic treatment within continuum medium electrostatics is possible. We can thus connect the leading eigenvalue to \textit{macroscopic} properties\cite{rosner_wannier_2015}. The other eigenvalues $v_{2/3}$ are assumed to be constant \cite{schonhoff_interplay_2016} and obtained by averaging over the ab-initio values. The macroscopic eigenvalue is fitted with
        \begin{align}
         v_1(\vec{q}) = \frac{3e^2}{2\varepsilon_0 A} \frac{1}{q (1 + \gamma q)}, \label{eq:fit_bare_coulomb}
        \end{align}
        where we use the area of the 2D hexagonal unit cell $A = \frac{\sqrt{3}}{2} a^2$ and the form factor $\gamma$ which describes how the effective height of the orbitals influences short wavelengths. 
          
        The screened Coulomb matrix $W(\vec{q})$ is assumed to have the same eigenbasis as $v(\vec{q})$ so that we can define its eigenvalues via the eigenvalues of the bare interaction and the dielectric function:
        \begin{align}
            W_i(\vec{q}) = \frac{v_i (\vec{q})}{\varepsilon_i (\vec{q})}. \label{eq:partially_screened_func}
        \end{align}
        The (leading) macroscopic eigenvalue of the dielectric function is given by\cite{rosner_wannier_2015, Emelyanenko2008}
        \begin{align}
			\varepsilon_1 (\vec{q},\omega) = \varepsilon_\infty \frac{1-\beta_1 \beta_2 \e^{-2qd}}{1 + (\beta_1 + \beta_2)\e^{-qd} + \beta_1 \beta_2 \e^{-2qd}}. \label{eq:epsilon}
		\end{align}
		with
		\begin{align}
			\beta_i  = \frac{\varepsilon_\infty - \varepsilon_{\t{env},i}(\vec{q},\omega)}{\varepsilon_\infty + \varepsilon_{\t{env},i}(\vec{q},\omega)},
		\end{align}
		which includes the dielectric functions of the material above and beneath the monolayer $\varepsilon_{\text{env},i}(\vec{q},\omega)$. 
		The microscopic  screening effects described by $\varepsilon_{2/3}$ are again assumed to be momentum independent, i.e. local, constants.
		Thus, by fitting all $\varepsilon_i$ to the ab intio values for the free standing monolayers (setting $\varepsilon_{\text{env},i}(\vec{q},\omega) = 1$) we gain fully analytic and material-realistic models for the Coulomb interaction matrix elements $W^\text{TMDC}$ in the orbital basis. The corresponding fitting parameters are given in Tab.~\ref{table:fit_parms}.
		
		By modifying $\varepsilon_{\text{env},i}(\vec{q},\omega)$ we can additionally include external screening effects from some material below or above the monolayer yielding analytic descriptions of $W^\text{TMDC}_\text{env}$. In the main text we mostly consider a single substrate, i.e., we set $\varepsilon_{\text{env},1} (\vec{q},\omega) = \varepsilon_\t{sub} (\vec{q},\omega)$ and $\varepsilon_{\text{env},2}(\vec{q},\omega) = 1$. In Section \ref{sec:results_static} we use a dielectric constant $\varepsilon_\text{sub} (\vec{q},\omega) = \varepsilon_\t{sub}$ and in Section \ref{sec:results_dynamic} we incorporate the frequency dependence via the plasmon-pole approximation and set $\varepsilon_\text{sub} (\vec{q},\omega) = \varepsilon_\text{sub} (\omega)$ according to Eq.~\eqref{eq:ppm}.
  
	 	Additionally, the WFCE approach allows us to correct the artificially introduced self-interaction effects within the super-cell setup used in the ab initio calculations.
	 	To do so, we performed RPA calculations for freestanding monolayer for different vacuum heights $h_\t{vac}$ between $15$\,\AA \ and $40$\, \AA \ and extrapolated the results to infinite vacuum heights
		\begin{align}
			W_{\alpha \beta} (\vec{q}, h_\t{vac} ) = W_{\alpha \beta} (\vec{q}, \infty ) + \frac{b_{\alpha \beta}(\vec{q})}{h_\t{vac}}.
		\end{align}
		The fitting parameter listed in Tab.~\ref{table:fit_parms} results from fits to these extrapolated Coulomb interaction matrix elements.

\newpage

\section{Substrate dielectric constants}
We present macroscopic dielectric constants for a few typical substrate materials in Table ~\ref{appendix:table_constant}.

\begin{table}[hpt]
\centering
\begin{tabular}{l||l|l|l|l|l}
substrate     & SiO$_2$                                 & HfO$_2$ & Si                         & GaAs                       & hBN                                  \\ \hline
$\varepsilon_\infty$ & $\approx 3.6$                               & $25$      & $\approx 12$                   & $\approx 13$                   & $\approx~(1.8 - 3.3) $                 \\
Reference     & [\onlinecite{Robertson2004_high_dielectric}] &   [\onlinecite{Robertson2004_high_dielectric}]      & [\onlinecite{Sze2006_constants}] & [\onlinecite{Sze2006_constants}] &  [\onlinecite{Wang2016,Laturia2018}]
\end{tabular}
\caption{Static dielectric constants for a few typical substrate for 2D materials} \label{appendix:table_constant}
\end{table}

\section{$G \Delta W$ convergence} \label{appendix:gdw_convergence}

\begin{figure}[htb]
	\includegraphics[width=\columnwidth]{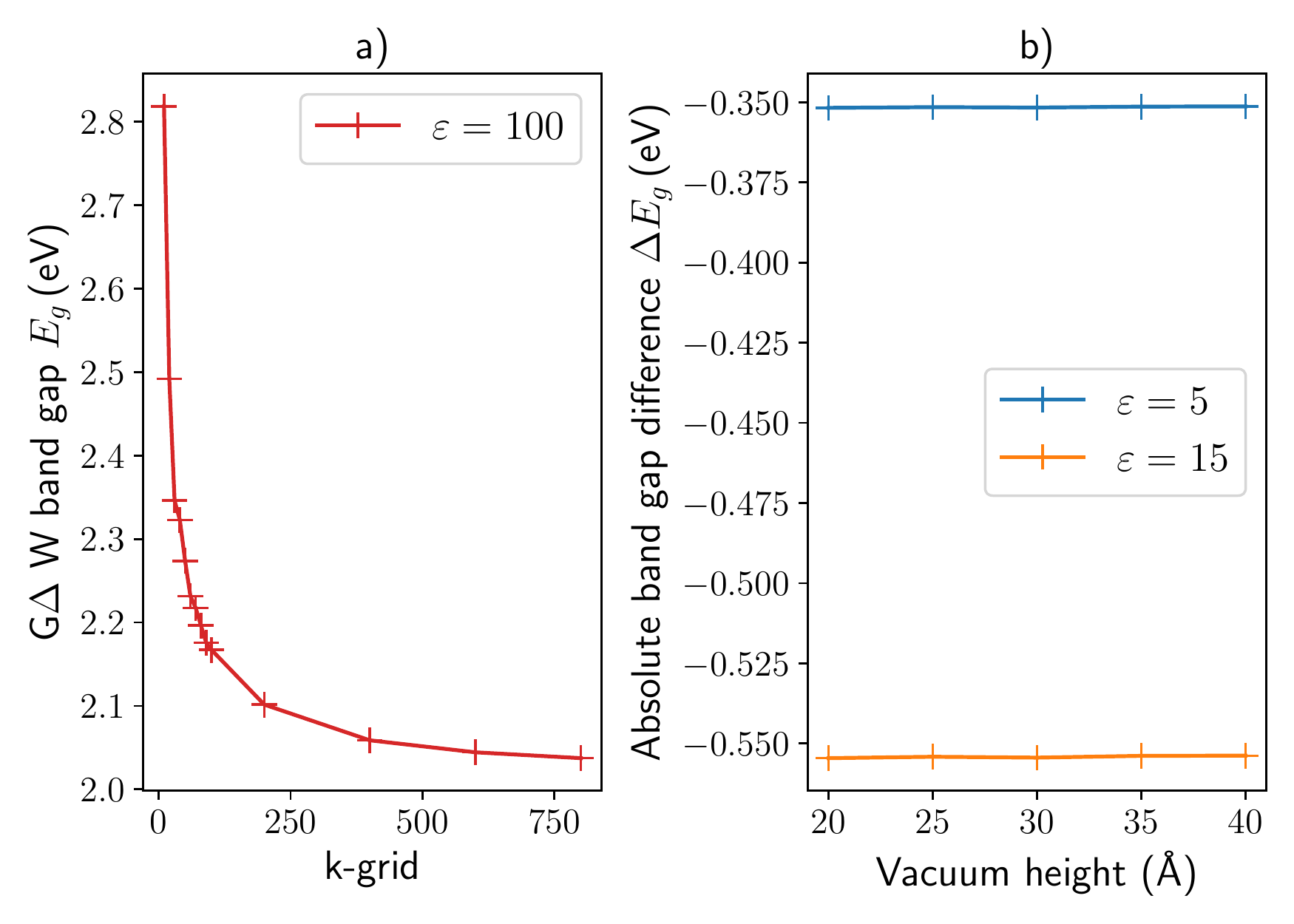}
	\caption{(a) Convergence of the $G \Delta W$ band gap for $\varepsilon_\t{sub} = 100$ and WS$_2$ depending on the k-grid  and (b) convergence of the absolute band gap difference depending on the vacuum height included in G$_0$W$_0$ calculations.  } \label{fig:GdW_convergence}
\end{figure}

In Fig.~\ref{fig:GdW_convergence} (a) we show the WS$_2$ band gap for a substrate dielectric constant of $\varepsilon_\text{sub} = 100$ as a function of the k-grid as obtained from the $G \Delta W$ approach. Due to the strongly peaked form of $\Delta W(\vec{q})$ in momentum space, rather fine k-meshes are needed to converge these $G \Delta W$ calculations. We use for all static calculations $400 \times 400$ k-points resulting in band gap inaccuracies smaller than $0.02\,$eV. For dynamic calculations, we use $100 \times 100$ k-points and $\omega$ grids from $-30$\, eV to $30$\, eV with 600 points.

In Fig.~\ref{fig:GdW_convergence} (b) we show the dependence of the absolute band gap changes for WS$_2$ for different dielectric constants on the vacuum height of the underlying $G_0W_0$ calculation for the freestanding monolayer. In these $G_0W_0$ calculations the quasi-particle band gap is underestimated but slowly converges with larger vacuum height. However, we see nearly no influence on the \textit{absolute} band gap changes thus we chose $c=20$\,\AA \ for all investigated TMDCs.

\section{Image Charge Effects to the Spatial Extent of the Band Gap Modulations in Coulomb Engineered Heterostructures} \label{appendix:HS_IC}

At any dielectric interface image charge effects arise if there are source charges on at least one side. 
The image charges alter the original potential created by the source charges. 
Hence, the Coulomb interaction within the homogeneous monolayer will be altered by the dielectric interfaces in the environment. 
In the heterojunctions depicted in Figs.~\ref{fig-1}\,(c) and \ref{fig-6}, there are vertical and lateral dielectric interfaces.
Here, we give an estimate of how image charges associated with the lateral interface will affect the length scale of the band gap modulations in the Coulomb engineered heterostructure. 
Therefore, we calculate how the Coulomb interaction is affected by the lateral dielectric interface. 
From the analysis of the self-energy in the main text we know just the local and nearest-neighbour interaction terms give the most significant contributions to the self-energy. 
Since the nearest-neighbour interaction will be altered most by image charge effects associated with the lateral interface, we focus on this element, here.

\begin{figure}[h]
	\includegraphics[width=\columnwidth]{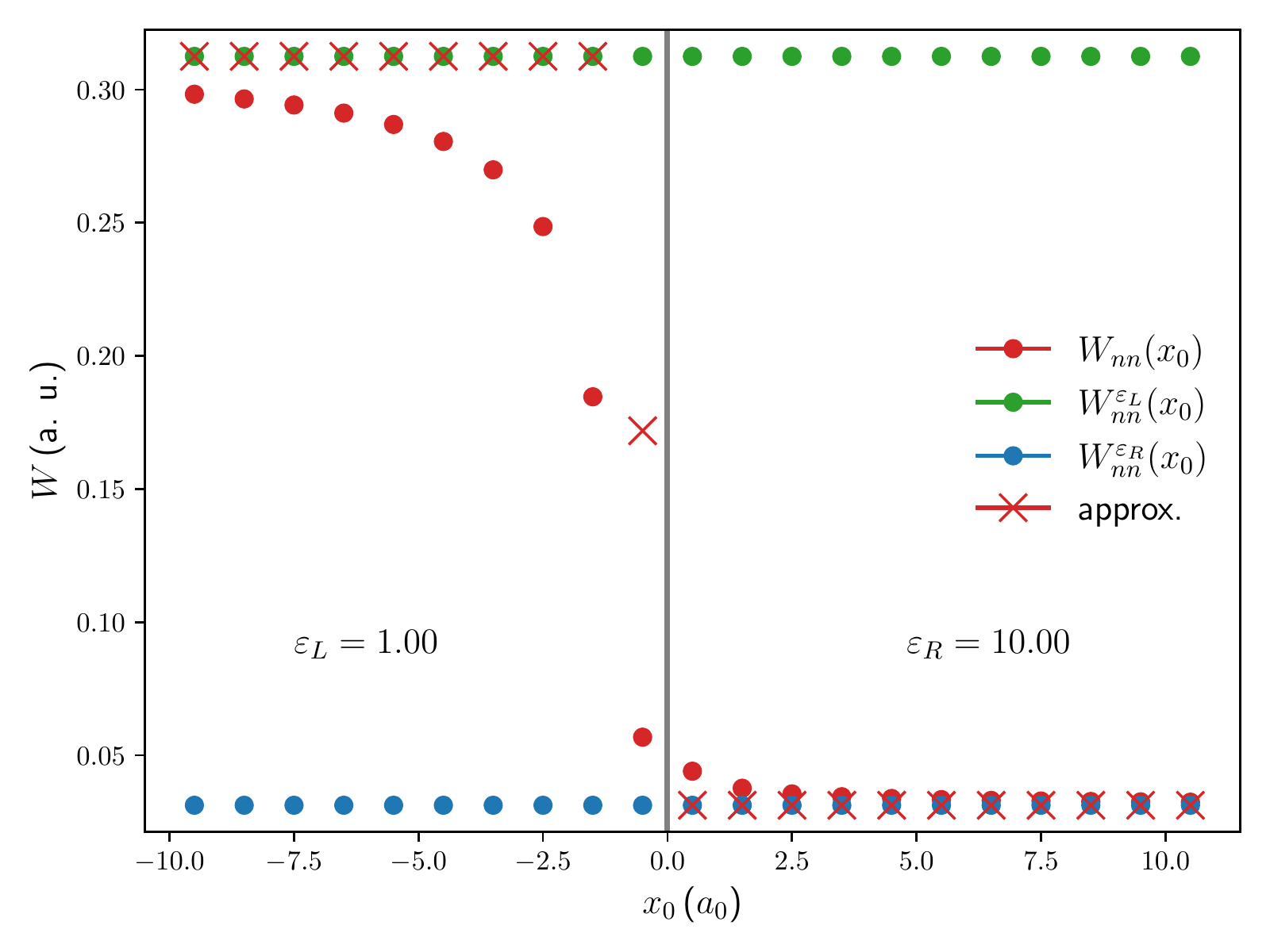}
	\caption{Lateral image charge effects to the nearest-neighbour Coulomb interaction. Red dots correspond to $W_{nn}(x_0)$ with $\varepsilon_L=1$ and $\varepsilon_R=10$. Green and and blue dots show $W_{nn}(x_0)$ without any dielectric interfaces for homogeneous $\varepsilon=1$ and $\varepsilon=10$, respectively. Red crosses depict the approximate values used in the main text. } \label{fig:HS_IC}
\end{figure}

To this end, we turn to the zero-height limit of the layer [i.e. $d \rightarrow 0$ in Eq.~(\ref{eq:epsilon})], where we can solve the dielectric problem analytically. 
The potential due to a source charge $q$ at $x_0<0$ on the left side of the interface (with dielectric constant $\varepsilon_L$) evaluated at $x<0$ and $y=z=0$ reads:
\begin{align}
	\Phi_{\varepsilon_L | \varepsilon_R}^{x \leq 0}(x_0, x) = 
	\frac{1}{\varepsilon_L} 
	\left[ \frac{q}{| x-x_0 |} + \frac{\tilde{q}}{| x+x_0 |} \right].
\end{align}
It contains the effect of the lateral interface through the image charge $\tilde{q} = q\frac{\varepsilon_L - \varepsilon_R}{\varepsilon_L + \varepsilon_R}$ on the right side (with $\varepsilon_R$ at $x>0$) at $x=-x_0$. 
From this, we can estimate the nearest-neighbour Coulomb interaction,
\begin{align}
	W_{nn}(x_0) = q \, \Phi_{\varepsilon_L | \varepsilon_R}^{x \leq 0}(x_0, x_0+a_0),
\end{align}
for our lattice model by setting $x_0$ to some lattice position and $x = x_0 + a_0 < 0$ to a corresponding nearest-neighbour distance on the $\varepsilon_L$ side (a similar expression can be dervied for $x = x_0 + a_0 > 0$). 
From this, we recover the limit of a laterally homogeneous dielectric environment for $|x_0|\to\infty$:
\begin{align}
	W_{nn}(|x_0|\to\infty) = \frac{1}{\varepsilon_L} \frac{q^2}{a_0}
\end{align}
At finite distance $x_0$ from the lateral interface, we find a dependence of the nearest-neighbor interaction strength $W_{nn}(x_0)$ on $x_0$ as depicted in Fig.~\ref{fig:HS_IC}.  The interface simulated in Fig.~\ref{fig:HS_IC} with $\varepsilon_L=1$ and $\varepsilon_R=10$ has same dilectric contrast as the example shown in Fig.~\ref{fig-6}.
In detail, we see that that $W_{nn}(x_0)$ smoothly interpolates the two homogeneous limits, reflecting the modulation of the Coulomb potential due to the lateral interface.
Within about $5$ unit cells away from the interface $W_{nn}(x_0)$ has closely approached the respective homogeneous ($|x_0|\to \infty$) limit. 
From the comparison to the approximation we used in the main text (here indicated by the red crosses), we thus see that taking these lateral image-charge effects into account will slightly smoothen the band gap modulations at the interface. 
However, the band gap modulation will still take place within about $6$ to $7$ lattice constants.

\bibliography{refs}

\end{document}